# Basic Connection between Superconductivity and Superfluidity

Mario Rabinowitz
Electric Power Research Institute
and
Armor Research
715 Lakemead Way, Redwood City, CA 94062-3922
E-mail: Mario715@earthlink.net

## Abstract

A basic and inherently simple connection is shown to exist between superconductivity and superfluidity. It is shown here that the author's previously derived general equation which agrees well with the superconducting transition temperatures for the heavy-electron superconductors, metallic superconductors, oxide superconductors, metallic hydrogen, and neutron stars, also works well for the superfluid transition temperature of 2.6 mK for liquid $^3$He. Reasonable estimates are made from $10^{-3}$ K to $10^9$K -- a range of 12 orders of magnitude. The same paradigm applies to the superfluid transition temperature of liquid $^4$He, but results in a slightly different equation. The superfluid transition temperature for dilute solutions of $^3$He in superfluid $^4$He is estimated to be ~ 1 to 10μK. This paradigm works well in detail for metallic, cuprate, and organic superconductors.

## 1. INTRODUCTION

The experimental discovery in 1972 of superfluidity in liquid $^3$He (L$^3$He) at 2.6mK was long preceded by predictions of the critical temperature $T_c$ for this transition at $\lesssim$ 0.1K which was then well below the lower limit of the known experimental data. These predictions were based on the BCS theory (Bardeen, Cooper, and Schrieffer, 1957) where the sensitive exponential dependence of $T_c$ makes it hard to make accurate predictions of $T_c$. Theoretical papers (Pitaevski ,1959; Brueckner et al, 1960; Emery and Sessler, 1960) incorporated pairing of $^3$He Fermi atoms to make Bosons by analogy with the Cooper pairing of Fermi electrons in metals. The theoretical and experimental $^3$He work are thoroughly discussed with ample references in excellent review papers respectively by Leggett (1975) and Wheatley (1975). Betts' (1969) excellent tutorial-review paper covers both superconductivity and only the superfluidity of $^4$He, as superfluidity had not yet been discovered in $^3$He. However, Betts does discuss dilute solutions of $^3$He in $^4$He as the best and then most recently realized examples of Fermi degeneracy.

When the experimental work showed that $T_c$ must be well below the theoretical prediction of 0.1K, theoreticians pushed $T_c \sim 10^{-9}$ K -- then beyond the hope of experimental verification. Agreement of the theory with experiment was established after the experimental detection by Osheroff, Richardson, and Lee (1972a) of remarkable features $\leq$ 3 mK of the pressurization curve (pressure vs time) of $L^3He$ in equilibrium with solid $^3He$ at about 34.4 bar. Originally, they interpreted this in terms of effects in the solid $^3He$. Their subsequent NMR experiments established that the effects were in the $L^3He$ (Osheroff et al, 1972b) with $T_c \leq$ 3 mK for the superfluid transition.

## 2. ANALYSIS FOR $^3He$

As the temperature is decreased in a standard Bose-Einstein (B-E) gas, particles should start a B-E condensation (transition) into a ground state, i.e. the superfluid state when the thermal wavelength $\lambda_T$ is comparable to the interparticle spacing d: $\lambda_T = h/[2\pi mkT_c]^{1/2} \sim d$. Here h is Planck's constant, m is the Boson mass, k is the Boltzmann constant, and $T_c$ is the critical (transition) temperature. I prefer the de Broglie wavelength $\lambda$ rather than $\lambda_T$ for reasons of clarity and comparison with my previous work (Rabinowitz, 1987, 1988, 1989a, 1989b, and 1990). This is also a consideration because of the close proximity of the carriers in the B-E condensation of $L^3He$ and of $L^4He$. As shown in Figure 1, $(1/2)\lambda$ encompasses the centers of mass of two pairs of Fermions when

$$\lambda > = 4d = 4n_s^{-1/3} . \qquad (1)$$

$n_s$ is the number density of particle pairs which can have an effective interaction for condensation because they have energies within $kT_c$ of the Fermi surface.

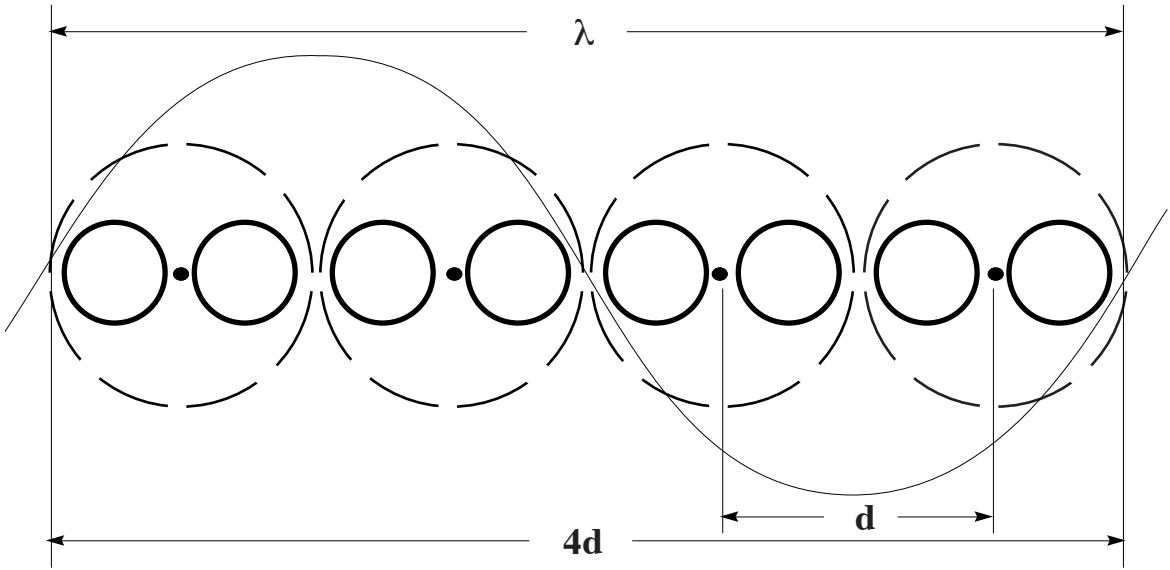

**Fig. 1.** $^3$He Fermion pairs.

$$n_s \sim \frac{kT_c}{E_F}\, n\,, \qquad (2)$$

where n is the number density of Fermions ($^3$He in the particular case we are considering) in the normal state, and $E_F$ is the three-dimensional Fermi energy.

$$E_F = \frac{h^2}{2m_e}\left(\frac{3n}{8\pi}\right)^{2/3} \approx \frac{h^2}{8m_e}n^{2/3}\,, \qquad (3)$$

where $m_e$ is the effective mass of the Fermion. In the case of $^3$He, the effective mass $m_e$ may be due to a screening cloud of additional atoms which act like a quasiparticle. This is not unlike the classical effective mass of a body moving through a liquid, which also moves some of the liquid along with it. The quasiparticles fill the Fermi sea up to the Fermi momentum.

For a particle pair of momentum p and mass $2m_e$

$$\lambda = \frac{h}{p} = \frac{h}{\left[2(2m_e)\left(\frac{1}{2}kT_c\right)\right]^{1/2}} , \tag{4}$$

where f is the number of degrees of freedom per particle pair. For three-dimensional $^3$He, we will simply take f =3. See Rabinowitz 1989a for a more general discussion of f.

Combining equations (1) through (4), we find the transition temperature

$$T_{cF} = \frac{h^2 n^{2/3}}{(8f)^3 m_e k} . \tag{5}$$

In general, equation (5) applies to paired Fermion particles forming a Boson gas, be they electrons or atoms. This is the expression derived as equation (5) in Rabinowitz 1989a. However, due to a misprint the $(8f)^3$ was printed incorrectly as $8f^3$. Equation (5) gave a reasonable estimate for the transition temperatures of heavy-electron superconductors, metallic superconductors, oxide superconductors, metallic hydrogen, and neutron stars. It thus did a good job in covering a range of 9 orders of magnitude from $\sim 0.5$K to $10^9$ K. For three-dimensional L$^3$He, f =3. (As the dimension-ality is lowered to quasi-2 or quasi-1 dimensions, the effective mass may change.) The effective mass and spin degeneracy should be used in conjunction with a more precise derivation. For now the free mass will suffice.

We shall now see that equation (5) also works well for the superfluid transition temperature of liquid $^3$He. We need only two parameters, which are experimentally determined. Atkins (1959) gives the density of L$^3$He as $0.082$ gm/cm$^3$ at T < 1 K. This implies that the number density of $L^3He$ atoms is n $\approx 1.6 \times 10^{22}/cm^3$. Taking $m_e \approx m_{3He} \approx 5 \times 10^{-24}$ gm, equation (5) gives

$T_{cF} \approx 2.8$ mK

in excellent agreement with the experimental value of 2.6 mK at $\approx$ 30 bar pressure as given by Wheatley (1975). $T_c$ decreases by a small factor with decreasing pressure, and there are two superfluid phases in $L^3He$.

In the $^3He$-A phase, the nuclear spins of the two $^3He$ atoms are parallel to each other and tend to be perpendicular to the axis of orbital motion. In the $^3He$-B phase the correlation is more abstruse. There are regions of pressure and temperature where each phase exists separately. Because all the pairs of a given kind are in the same state, the spin and orbital motion correlations of each kind exist throughout the superfluid as a whole. Thus, unlike superfluid $^4He$ which has 0 spin and is thus insensitive to perturbations which allow it to remain in the superfluid state, superfluid $^3He$ has orientation which can be affected by external factors such as flow motion, interaction with surfaces, and applied magnetic and electric fields (Wheatley, 1975; Leggett, 1975).

### 3. COMPARISON BETWEEN $L^3He$ AND SUPERCONDUCTORS

The creation of superfluidity by the pairing of $^3He$ Fermi atoms to make Bosons is analogous to the Cooper pairing of Fermi electrons in metals to create superconductivity. However, there is one important difference. In metallic superconductors, the paired electrons form a singlet state of zero spin (relative angular momentum), whereas the strong hard-core repulsion of $^3$He atoms causes them to pair in a triplet state of spin 1.  In the heavy Fermion superconductors, it is possible that the electrons also pair in a triplet state of spin 1.

It is interesting to compare $kT_c/E_F$ for L$^3$He and the different classes of superconductors as this is a measure of the relative participation of potential carriers near $T_c$.  Let us look at them in increasing order for $T_c$.  For L$^3$He:  $kT_c/E_F \sim 22$ μeV/44 meV $\sim 5$ x $10^{-4}$.  For the heavy Fermion superconductors:  $kT_c/E_F \sim 0.1$meV/1eV $\sim 10^{-4}$.  For the metallic superconductors:  $kT_c/E_F \sim 1$ meV/10 eV $\sim 10^{-4}$.  For the ceramic oxides: $kT_c/E_F \sim 10$ meV/2 eV $\sim 5$ x $10^{-3}$.  For metallic H it is expected that:  $kT_c/E_F \sim 20$ meV/4 eV $\sim 5$ x $10^{-3}$.  For a neutron star it is expected that:  $kT_c/E_F \sim 0.1$MeV/200 MeV $\sim 5$ x $10^{-3}$.  To first order it is remarkable how close all of these diverse quantum fluids are in their values of   $kT_c/E_F$.

## 4.  PREDICTIONS FOR L$^3$He IN SUPERFLUID $^4$He

It is difficult to observe superfluidity in most substances because they go into the solid state before the extremely low temperatures are reached at which they might become superfluids.   However, the combination of  $^3$He in superfluid $^4$He does not have this problem, and allows for a prediction in a novel system for which superfluidity has not yet been observed.

According to my paradigm, if the temperature is lowered sufficiently a dilute solution of $^3$He atoms in superfluid $^4$He should become a quantum fluid when the $^3$He quantum wavelength is more than 4 times the $^3$He interparticle spacing. This is analogous to the pairing of electrons in a metallic superconductor where the coherence length is quite large, $\sim 10^4$ A for a pure superconductor. The pairing of $^3$He atoms in L$^3$He is more analogous to the pairing of electrons in a ceramic oxide superconductor where the coherence length is quite small, $\sim$ few Angstoms. At close distances (high concentration), because of the hard core repulsion, only the triplet state of parallel spins seems possible for $^3$He. However, at low concentrations with large distances between the atoms, the singlet state of 0 spin (antiparallel spins for the two atoms) would be energetically preferred. At intermediate concentrations, both singlet and triplet pairs may form. With the singlet state, and $^3$He-A and $^3$He-B for the triplet state, 7 combinations would be possible for this 3-component superfluid.

As an analog to the metal lattice, $^4$He has the advantage not only of not becoming solid (which would impede motion of $^3$He) at these extremely low temperatures, but also of being a superfluid itself. The $^4$He environment may act like a vacuum with negligible perturbation on the $^3$He. It may even be the analogue of a solid superconductor in which paired electrons move like a charged superfluid. However, $^3$He in $^4$He interactions may be more important as there may be many strong hard core interactions in the scattering of the $^3$He atoms in the $^4$He environment. These collisions may reduce the coherence and hence $T_c$. In a way superfluid $^4$He may to first approximation act like a kind of vacuum with mass, with respect to $^3$He atoms. This is not too unlike a contemplation by Finkelstein (1988, 1989) regarding the properties of space-time. The two-fluid (normal and superstates) interpenetrating model of superconductivity and of superfluidity works quite well. Rodriguez-Nunez and Tello-Llanos (1991) have recently developed a two-fluid model using the Dirac formalism.

Only dilute solutions of up to 10% $^3$He in $^4$He are possible. We shall here concern ourselves only with th volume displacement due to foreign particles -- $^4$He in this case. A sizable correction must be made for a volume reduction effect due to the presence of $^4$He. The corrected number density $\eta$ of $^3$He in $^4$He is

$$\eta = \frac{N_3}{V_3} = \frac{gN}{g'V} = \frac{g}{g'}n \approx n$$

where n is the overall number density, $N_3$ is the number of $^3$He atoms, and $V_3$ is the volume available to them. For simplicity, I assume that the volume fraction g available to the $^3$He atoms is approximately the same as the number fraction g'. This is a good approximation, as the effective volumes of $^3$He and $^4$He are close enough within the other approximations of my calculations. One could easily carry g and g' through in the analysis.

Thus, in this case the Fermi energy is still given by equation (3), since the corrected number density $\eta$ of $^3$He results to first approximation in the overall number density n. However, $\lambda$ is related to the number density $n_3$ of $^3$He , so that equation (1) becomes

$$\lambda \geq 4d = 4n_s^{-1/3} = 4\left(\frac{kT_c}{E_F}n_3\right)^{-1/3} = 4\left(\frac{kT_c}{E_F}gn\right)^{-1/3} . \qquad (6)$$

Combining equation (6) with (2)-(4) and , we get the superfluid transition temperature for a dilute Fermi liquid,

$$T_{dF} \leq \frac{h^2 n^{2/3}g^2}{(8f)^3 m_e k} . \qquad (7)$$

For a 5% concentration of $^3$He atoms in L$^4$He, equation (7) predicts $T_{dF}$ ~ $10^{-5}$K for $m_e \approx m_{3He} \approx 5 \times 10^{-24}$ gm. If $m_e \sim 10\, m_{3He}$ for a quasiparticle of one $^3$He surrounded by $^4$He, $T_{dF}$ ~ $10^{-6}$K. This is significantly higher than more elaborate theories that put this transition at ~ $10^{-9}$ K. Experiments with dilute solutions are under way, but no superfluid transition has yet been observed for the solute $^3$He atoms.

## 5. ANALYSIS FOR $^4$He

It will next be shown that the same paradigm applies to the superfluid transition temperature of liquid $^4$He (L$^4$He), but results in a slightly different equation that also agrees well with experiment. In L$^4$He, pairing of the atoms is not necessary as these particles are already Bosons, and there is no Fermi energy below which all the states are filled for Fermions because of the Pauli principle. Thus all the particles may participate in a B-E condensation, not just a fraction $kT_c/E_F$ as in the case of Fermions.

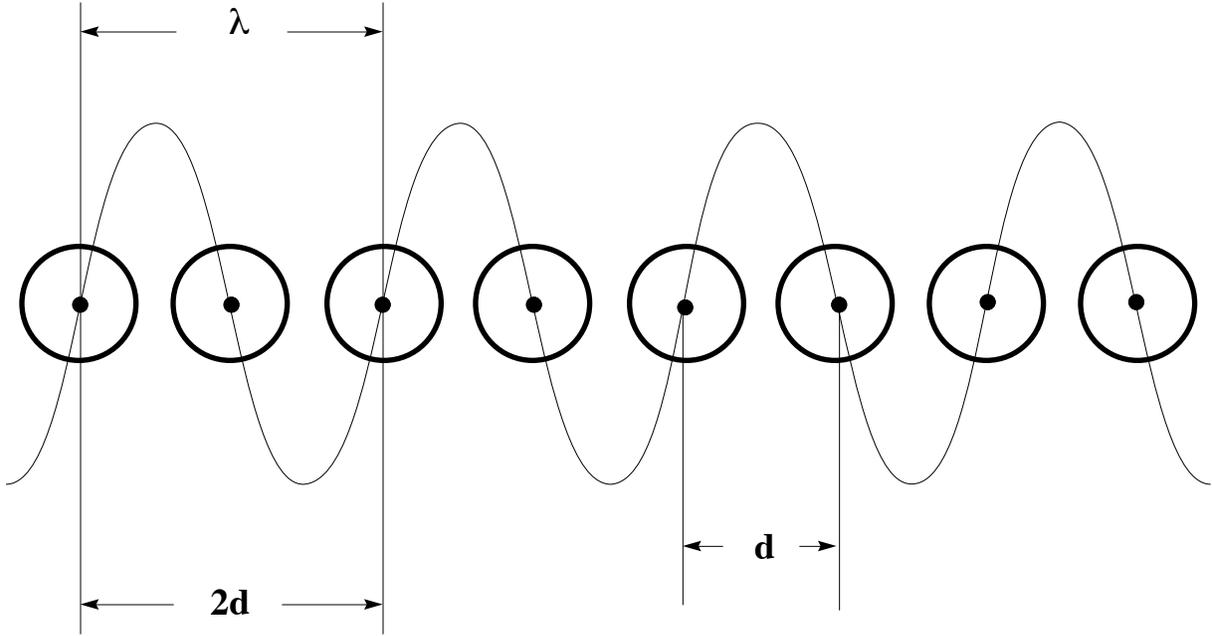

**Fig. 2.** $^4$He Bosons.

As shown in Figure 2, since pairing does not occur, $(1/2)\lambda$ encompasses the centers of two Bosons when

$$\lambda \geq 2d = 2n^{-1/3} , \tag{8}$$

where n is the number density of Bosons ($^4$He in this case).

$$\lambda = \frac{h}{p} = \frac{h}{\left[2(2m_e)\left(\frac{1}{2}kT_c\right)\right]^{1/2}} , \tag{9}$$

where the $2m_e$ in equation (4) has been replaced by $m_e$ since pairing is not necessary.

Combining equations (8) and (9) we find the superfluid transition temperature for a Boson gas is

$$T_{cB} = \frac{h^2 n^{2/3}}{4fm_e k} \tag{10}$$

For three-dimensional L$^4$He, f =3. Atkins (1959) gives the density of L$^4$He as 0.1465 gm/cm$^3$ at T ≤ 2.17 K (λ point). Thus n ≈ 2.2 x 10$^{22}$/cm$^3$. Taking $m_e$ ≈ $m_{4He}$ ≈ 6.7 x 10$^{-24}$ gm, equation (10) gives

$$T_{cB} ≈ 3 \text{ K}$$

in good agreement with the experimental value of 2.17 K for the superfluid transition of L$^4$He (Atkins 1959).

It is interesting to note that for f =3, equation (10) gives

$$T_{cB} = \frac{h^2 n^{2/3}}{12 m_e k} \qquad .$$

This is almost identically the same as the $T_c^{\textbf{BE}}$ obtained from a more rigorous, but also much more difficult derivation as found in textbooks (Huang, 1963) for a B-E condensation for pure Bosons:

$$T_c^{\textbf{BE}} = \frac{h^2 n^{2/3}}{2\pi m k [\zeta(\frac{3}{2})]^{2/3}} = \frac{h^2 n^{2/3}}{2\pi m k [2.612]^{2/3}} = \frac{h^2 n^{2/3}}{11.92 \, m k} \qquad , \qquad (11)$$

where $\zeta(x)$ is the Riemann zeta function of x, and $\zeta(3/2) = 2.612...$ . The ratio $T_c^{\textbf{BE}}/T_{cB} = 12/11.92 = 1.007$.

## 6. DISCUSSION

It is remarkable how well this simple paradigm works over a range of 12 orders of magnitude from 10$^{-3}$ K to 10$^9$K. It may seem surprising that it works so well, since an interaction between the particles is not part of the paradigm. Perhaps this should not be so surprising. The B-E condensation temperature $T_c^{\textbf{BE}}$ for $^4$He works quite well, and it does not include an interaction. Equation (11) gives $T_c^{\textbf{BE}}$ = 3.1 K, whereas the experimental value is $T_C$ = 2.17 K. The small difference can be attributed to an interaction potential. Possibly it could be attributed to an effective mass $m_e >$ $m_{4_{He}}$ . In any event only a small correction is necessary. It may well be that there are large interactions, but that their net effect is small.

For each class of materials, interactions can reduce $T_c$ from the ideal values given by my paradigm so that it basically gives an upper limit of $T_c$ for that class. The quantum condition appears to be the overriding effect in the sense of a general principle that determines the upper limit $T_c$ with little need for microscopic calculations. Three readily stated examples will illustrate my point.

1. If we wish to find the center of mass of the exhaust gases from a rocket ship in space we can either do a tedious calculation of the trajectories and momentum transfer of all the $>> 10^{27}$ particles, or we can simply apply the principle of the conservation of the center of mass in following the motion of the rocket ship.

2. To determine the radiation from a uniformly charged sphere oscillating radially, we could microscopically calculate the radiation from each point on the sphere taking interference into account. Or we could simply note that the center of charge is not accelerating, and hence there is no radiation. Done correctly, either approach will give the right answer; but one way is tremendously more difficult than the other.

3. The diffusion time constant for a magnetic field into a medium of permeability $\mu$ and conductivity $\sigma$ is $\tau = \dfrac{\mu\sigma\delta^2}{2}$ . Where $\delta$ is the penetration depth. We could make laborious measurements of $\mu$, $\sigma$, and $\delta$ to determine $\tau$. On the other hand knowing that $\delta = \left[\dfrac{2}{\mu\sigma\omega}\right]^{1/2}$ ($\omega$ is the angular frequency), let's us find $\tau$ quite easily: $\tau = 1/\omega$.

If the BCS paradigm applies to dilute $^3$He, then scattering from $^4$He should have no effect on the center of mass motion of a $^3$He pair. BCS assumes that each member of a pair has equal and opposite momentum. This guarantees that as one particle is scattered, the other one will move in the opposite direction to conserve the center of mass motion unimpeded.

There may be a hysteretic effect for $T_c$ for some superconductors or superfluids by preventing the pairing of Fermions. Apply a large enough magnetic field to keep the Fermions unpaired as T is lowered to near 0 K. It is commonly expected that the super state will return upon removal of the magnetic field. However, the Pauli exclusion principle may inhibit pairing, and the normal unpaired state may remain as a Fermi fluid -- cf. eq.(2). Direct cooling without a magnetic field leads to the paired super state.

## 7. CONCLUSION

Good experiments will hopefully answer the many questions raised in this paper. Past theory in the prediction of $T_c$ in both the arenas of superconductivity and superfluidity has not been too successful. To my knowlege, the $T_c$ of any superconductor or superfluid has rarely been correctly predicted in advance of the measurement. This is in part due to the inherent complexity and difficulty of the previous approaches.

I have great respect for the extant theories of superconductivity and superfluidity. They are great intellectual achievements. Their complexity does not make them erroneous. Neither does the simplicity of my paradigm make it wrong. Hopefully, the two approaches can balance each other. There is a need for both. A good theory must accurately portray a large class of observations with a model that has only a few parameters in it. Furthermore, it must make correct predictions regarding prospective findings.

My paradigm is a useful description of quantum fluids that fits the above requirements well. It is not designed to give a microscopic description. That is done well by the existing theories. It does show that the B-E condensation as a general feature common to a broad range of states of matter appears to be as important as any particular interaction mechanisms in the transition of both the superconducting and superfluid states from $10^{-3}$K to $10^9$K. When pairing occurs at $T \geq T_c$, then the interactions and the pairing mechanism may be irrelevant as the transition is then primarily limited by the condensation temperature.

## ACKNOWEDGMENT

I wish to express deep appreciation to Floyd Culler and David Gahagen for their valuable interest, encouragement, and support.

## REFERENCES

Atkins, K. R. (1959), *Liquid Helium,* Cambridge Press, Cambridge.

Bardeen, J., Cooper L. N., and Schrieffer, J. R. (1957). *Physical Review* **108**. 1175.

Betts, D. S. (1969). *Contemporary Physics* **10**, 241

Brueckner, K. A., Soda, T., Anderson, P. W., and Morel P. (1960). *Physical Review* **118**. 1442.

Emery, V. J., and Sessler A. M. (1960). *Physical Review* **119**. 43.

Finkelstein, D. (1988). *International Journal of Theoretical Physics,* **27**, 473.

Finkelstein, D. (1989). *International Journal of Theoretical Physics,* **28**, 1081.

Huang, K. (1963). *Statistical Mechanics*, Wiley, New York.

Leggett, A. J. (1975). *Reviews of Modern Physics* **47**, 331.

Osheroff, D. D., Richardson, R. C., and Lee, D. M. (1972a). *Physical Review Letters* **28**. 885.

Osheroff, D. D., Gully,W. J., Richardson, R. C., and Lee, D. M. (1972b). *Physical Review Letters* **29**. 920.

Pitaevskii, L. P. (1959). *Zh. Eksp.* Teor. Fiz. **37**, 1794. [(1960) Sov. Phys. JETP **10**, 1267.]

Rabinowitz, M. (1987). In *Proceedings: EPRI Workshop on High-Temperature Superconductivity.*

Rabinowitz, M. (1988). In *Proceedings: EPRI Conference on High-Temperature Superconductivity.*

Rabinowitz, M. (1989a). *International Journal of Theoretical Physics,* **28**, 137.

Rabinowitz, M. (1989b). *Physica,* **C162-164**, 249.

Rabinowitz, M. (1990). *Advances in Cryogenic Engineering*, **36A**, 21. Plenum Press, New York.

Rodriguez-Nunez, J. J., and Tello-Llanos, R. (1991) *International Journal of Theoretical Physics,* **30**, 857.

Wheatley, J. C. (1975). *Reviews of Modern Physics,* **47**, 415.